\documentclass[usenatbib,referee]{mn2e}
\usepackage{graphicx}
\title{The optical spectral slope variability of 17 blazars}
\author[Shao Ming Hu et al.]{Shao Ming Hu$^{1,2}$, G. Zhao$^{1}$\thanks{Corresponding author. E-mail: gzhao@bao.ac.cn}, H. Y. Guo$^{3}$,  X. Zhang$^{4}$ and Y. G. Zheng$^{4}$ \\
       $^{1}$National Astronomical Observatories, Chinese Academy of Sciences, A20 Datun Road, Beijing, 100012, China\\
       $^{2}$Department of Space Science and Applied Physics, Shandong University at Weihai, 180 Cultural West Road, \\ Weihai,
       Shangdong, 264209, China\\
       $^{3}$Computer Science and Technology College, Harbin Institute of Technology at Weihai, Weihai,
       Shangdong, China\\
       $^{4}$Department of Physics, Yunnan Normal University, Kunming, Yunnan, China}

\begin{document}
\date{Received date  / Accepted date}

\pagerange{\pageref{firstpage}--\pageref{lastpage}} \pubyear{2006}

\maketitle

\label{firstpage}

\begin{abstract}
Many quasi-simultaneous optical observations of 17 blazars were
obtained from the previous published papers over last 19 years in
order to investigate the spectral slope variability and understand
the radiation mechanism of blazars. The long period dereddened
optical spectral slopes were calculated in this paper. Our analysis
upon the average spectral slope distribution suggests that the
spectra of Flat Spectrum Radio Quasars (FSRQs) and High energy
peaked BL Lac objects (HBLs) are probably deformed by other
emission components. The average spectral slopes of Low energy
peaked BL Lac objects(LBLs), which scatter around 1.5, show
a good accordance with Synchrotron Self-Compton (SSC)
loss-dominated model. We present and discuss the variability
between spectral slope and optical luminosity. The spectra  of all
HBLs and LBLs get flatter when they turn brighter, while for FSRQs
this trend does not exist or may be in a reverse situation. This
phenomenon may imply that there is a thermal contribution to the
optical spectrum for FSRQs. For the FSRQ 1156+295, there is a hint
that the slope gets flatter at both the brightest and faintest
states. Our result shows that three subclasses locate in different
regions in the pattern of slope variability indicator versus the
average spectral slope. Furthermore we also delineate that
relativistic jet mechanism is supported by the significant
correlation between the optical Doppler factor and the average
spectral slope.
\end{abstract}


\section{Introduction}

Blazars are a subset of Active Galactic Nuclei (AGNs), which are
compact, flat spectrum radio sources with highly variable and
polarized nonthermal continuum emission ranging from radio up to
X-ray and often to $\gamma$-ray frequency (e.g.
\citealt{fugmann,bregman,sillanpaa,maraschi,urry}). Their observed
properties  are mainly considered to originate from a relativistic
beaming jet \citep{rees}, which is possibly powered and
accelerated by a rotating and accreting supermassive black hole.
There are two classes of blazars, BL Lac objects (BL Lacs) and
FSRQs, the former one has a featureless optical continuum, while
the latter has a great many strong and broad emission lines.
According to the peak frequency of synchrotron emission relative
to the peak flux, BL Lacs are classified as LBLs, whose peak is
located in infrared or optical band, and HBLs, whose peak is
located in the ultraviolet or X-ray frequency
\citep{padovani,giommi1999}. For HBLs, the broadband spectral
index $\alpha_{RX}$ is less than or equal to 0.75 \citep{urry} or
0.80 \citep{sambruna}, while for LBLs, it is greater than 0.75 or
0.80.

The studies of the spectral energy distribution (SED) of blazars
are very important for the understanding of the physical radiation
mechanism and to constrain the parameters involved in physical
model, and the studies are crucial in analyzing individual emission
components. Synchrotron inverse Compton emission model predicts
that the spectrum gets harder as the source turns brighter. Many
investigators found that the variation amplitude of blazars
at high frequencies was larger than that at low frequencies,
namely, the spectrum became flatter when the flux increased, while
the spectrum became steeper when the flux decreased (e.g.
\citealt{racine,gear,massano}). But \citet{ghosh} suggested that
it might not be always correct (e.g. \citealt{ramirez}).
\citet{damicis} reported the optical spectral slope
variability of 8 BL Lacs. They found that the spectral slope
became flatter when the source turned brighter for all 8 objects.
Four objects showed marked correlation between spectral
slopes and \emph{R} magnitudes. \citet{trevese} analyzed
 the spectral slope variability of 42 PG quasars.
They concluded that the spectral variability must be intrinsic of
the nuclear component. Their numerical simulation showed that hot
spots on the disk were able to represent the observed spectral
variability, while single host galaxy contribution or single
changes of accretion rate was insufficient to explain the slope
variability. \citet{vagnetti} analyzed the data
from \citet{damicis} under a simple synchrotron mode and the
thermal bump (TB) mode. They showed that the spectral variability,
even restricted to the optical band, could be used to set limits
on the relative contribution of the synchrotron component and the
thermal component. They deduced that the spectral variability of
BL Lacs differed from the variability of quasars. The 8 BL Lacs
they studied were all LBLs. \citet{ramirez} presented that the
spectrum of a FSRQ PKS 0736+017 became softer when it turned
brighter. The correlation between spectral slope and flux was
strong. Its variability relation between slope and flux was
reverse to the result that was mentioned above for LBLs.

In this paper we show and discuss the optical spectral slope
variability of 17 blazars, which include 9 LBLs, 4 HBLs and 4
FSRQs. In section 2 we describe the data and data reduction
process. The average optical spectral slope distribution is
analyzed in section 3, the spectral slope variability is given in
section 4, and conclusions are presented in the last section.

\section{Data and data reduction}

We collected optical observations from the published papers
in period of $1985\sim2003$, which were made in the standard
Johnson-Cousins \emph{B}, \emph{V}, \emph{R}, \emph{I} bands. The observations should be
simultaneous in all above-mentioned optical bands in order to get
the optical spectral slope, but it is difficult due to telescopes
and instruments. \citet{xie2005} reported that the minimum
variability timescales of 21 blazars were longer than 1 hour, and
most of them were longer than 2 or 3 hours. We regard this
quasi--simultaneous multi-band observations taken within 1 hour as
``simultaneous" observations (\citet{damicis} and \citet{fiorucci}
accepted quasi-simultaneous observations taken within 2 hours). So
we merely selected the quasi-simultaneous observations, which were
made in three(any three bands of \emph{B}, \emph{V}, \emph{R} and \emph{I}) or four bands(\emph{B},
\emph{V}, \emph{R}, \emph{I}) within 1 hour, as the raw data for accuracy. We selected a
group of quasi-simultaneous observations in three or four bands as
one data group, and most of the data groups were made within 30
minutes. Then the following reduction processes were applied to all
the selected raw data. All the observation data made in \emph{V} band
were corrected for the foreground Galactic interstellar reddening
and absorption with the employment of the extinction value
$A_{V}$, which is deduced from the maps of dust infrared emission
reported by \citet{schlegel}. The extinction values in other bands
were obtained by using the curve of \citet{cardelli}. The
reddening and absorption correction can be done by subtracting the
corresponding extinction value from the raw data. The conversion
from apparent magnitudes to optical flux densities was made by
using the zero magnitude equivalent flux density given by
\citet{mead}. The rest frequencies were obtained through the
effective wavelengths of the filters given by \citet{mead} and
the redshift given by \citet{veron}.

The radiation of blazars can be described by a single power law
$F_{\nu}=A\nu^{-\alpha}$ in optical band, where $\alpha$ is the
spectral slope. It was obtained by using the linear least
square fitting between $logF_{\nu}$  and $log\nu$. We can obtain
one $\alpha$ from each data group, but we accepted only the spectral
slope whose square of Pearson's linear correlation coefficient is
greater than 0.9 and whose standard deviation of fitting slope is
less than 0.4 (the typical standard deviation of the spectral
slope is less than 0.1). The aforesaid selection criterions
assure the accuracy of the slopes by cutting off the probably
wrong values due to quasi-simultaneous \citep{fiorucci}. After all
the above reduction, we accepted 1418 useful data groups from
all the raw data groups for 17 blazar samples whose data were
sufficient to investigate the spectral slope variability. Our main
results are listed in Table~\ref{table1}. The sequence columns in
Table~\ref{table1} are the source name; the redshift \emph{z} from
\citet{veron}; the average \emph{R} magnitude $\bar{R}$ and the variable
amplitude(max-min) of \emph{R} magnitude $M_{R}$; the average optical
spectral slope $\alpha_{ave}$ and the standard deviation of
$\alpha_{ave}$; the variable amplitude(max-min) of the optical
spectral slope $M_{\alpha}$; the slope of the linear regression
between $\alpha$ and the \emph{R} magnitudes, \emph{b}, (it is taken as the
slope variability indicator), followed by its standard deviation;
the correlation coefficient between $\alpha$ and \emph{R} magnitude,
$r^{2}$; the optical Doppler factor $\delta$; the reference of
$\delta$; the extinction value in \emph{V} band $A_{V}$; subclass of the
source; the number of useful data groups; the references of
optical photometric observations.

\begin{table*}
\begin{minipage}[t][]{\textwidth}
\caption{Principal data and results of 17 blazar samples}
\label{table1} \centering
\begin{tabular}{lcccccrlcclll}
\hline\hline
          (1)    &  (2)    & (3)mag              &(4)                        &(5)               &(6)               &(7)   &  (8)     &(9) &(10)            &(11)      &(12) & (13)\\
           name & z        &$\bar{R}(M_{R})$  &$\alpha_{ave}(\pm\sigma)$&$M_{\alpha}$      &b$(\pm\sigma)$ & $r^{2}$&$\delta$ & Ref. &$A_{V}$         &class     & N &Obs. Ref. \\
\hline    3C66A       & 0.444 & 14.40(1.07)   &1.18($\pm$0.27)          & 1.08       & 0.47($\pm$0.15)     &  0.61 & 2.99   & 1     &0.279           & LBL      &18 & 4,5,6,7,8   \\
          0235+164    & 0.940 & 15.83(1.52)   &2.62($\pm$0.36)          & 1.00       & 0.57($\pm$0.12)     &  0.86 & 6.50   & 1     &0.262           & LBL      &10 & 6,9         \\
          0323+022    & 0.147 & 16.33(1.01)   &0.99($\pm$0.43)          & 1.17       & 0.72($\pm$0.29)     &  0.54 & 1.54   & 2     &0.372           & HBL      &17 & 10,11       \\
          0422+004    & 0.310 & 14.52(2.16)   &1.26($\pm$0.23)          & 0.91       & 0.20($\pm$0.06)     &  0.48 & 1.57   & 3     &0.335           & LBL      &47 & 12,13       \\
          0716+714    & $>$0.30& 13.66(2.28)  &1.21($\pm$0.17)          & 1.16       & 0.21($\pm$0.01)     &  0.53 & 2.10   & 1     &0.102           & LBL      &749& 14,15,16,17    \\
          0735+178    & $>$0.42& 15.20(1.90)  &1.47($\pm$0.27)          & 1.10       & 0.27($\pm$0.08)     &  0.43 & 2.22   & 1     &0.116           & LBL      &54 & 6,7,18,19,20\\
          OI 090.4    & 0.266 & 15.13(1.17)   &1.33($\pm$0.20)          & 0.65       & 0.32($\pm$0.15)     &  0.49 & 3.56   & 1     &0.075           & LBL      &17 & 6,11,16,20,21 \\
          OJ 287      & 0.306 & 14.99(1.98)   &1.46($\pm$0.23)          & 1.09       & 0.16($\pm$0.06)     &  0.30 & 3.08   & 1     &0.094           & LBL      &77 & 6,7,8,20,22,23\\
          Mrk 421     & 0.031 & 12.79(0.26)   &1.18($\pm$0.24)          & 0.81       & 2.66($\pm$0.85)     &  0.77 & 1.40   & 1     &0.051           & HBL      &9  & 20,23       \\
          1156+295    & 0.729 & 16.23(3.21)   &1.10($\pm$0.23)          & 0.86       & -0.05($\pm$0.07)    & -0.18 & 1.59   & 1     &0.064           & FSRQ     &18 & 12,21       \\
          ON 231      & 0.102 & 14.04(1.52)   &1.25($\pm$0.15)          & 0.83       & 0.27($\pm$0.03)     &  0.56 & 1.35   & 1     &0.075           & LBL      &196& 24          \\
          3C 273      & 0.158 & 12.35(0.68)   &0.49($\pm$0.21)          & 0.76       & -0.66($\pm$0.13)    & -0.82 & 1.49   & 1     &0.068           & FSRQ     &15 & 11,12       \\
          3C 279      & 0.536 & 14.04(1.92)   &1.80($\pm$0.19)          & 0.67       & 0.20($\pm$0.12)     &  0.49 & 2.40   & 1     &0.095           & FSRQ     &10 & 25,26       \\
          1402+042    & 0.344 & 16.14(0.40)   &1.32($\pm$0.37)          & 1.44       & 0.92($\pm$1.04)     &  0.26 &$\cdots$&$\cdots$&0.086          & HBL      &13 & 6           \\
          2155-304    & 0.116 & 12.63(0.85)   &0.83($\pm$0.06)          & 0.18       & 0.16($\pm$0.04)     &  0.82 & 1.00   & 1     &0.071           & HBL      &12 & 27          \\
          BL Lac      & 0.069 & 14.08(2.14)   &1.66($\pm$0.26)          & 1.68       & 0.32($\pm$0.04)     &  0.62 & 2.43   & 1     &1.091           & LBL      &139& 7,18,20,28,29  \\
          3C 454.3    & 0.859 & 16.07(0.53)   &0.90($\pm$0.21)          & 0.80       & -0.45($\pm$0.15)    & -0.56 & 2.12   & 1     &0.355           & FSRQ     &17 & 12            \\
\hline\\
\end{tabular}
\flushleft{\texttt{Notes: $\cdots$ means no available data.\\
References---1. \citealt{xie2001}; 2.\citealt{xie1991}; 3.\citealt{lahteenmaki};
4.\citealt{xie1988a}; 5.\citealt{xie1988b}; 6.\citealt{xie1992};
7.\citealt{zhangx}; 8.\citealt{takalo1994}; 9.\citealt{takalo1998};
10.\citealt{villata2000}; 11.\citealt{fiorucci1996}; 12.\citealt{raiteri1998};
13.\citealt{massaro1996b}; 14.\citealt{qianbc2002}; 15.\citealt{ghisellini};
16.\citealt{bai1998};
17.\citealt{raiteri2003}; 18.\citealt{bai1999}; 19.\citealt{qianbc2004}; 20.\citealt{takalo1991};
21.\citealt{xie1994}; 22.\citealt{qianbc2003}; 23.\citealt{takalo1993}; 24.\citealt{tosti};
25.\citealt{xie2002}; 26.\citealt{villata1997};
27.\citealt{tommasi}; 28.\citealt{fan2001}; 29.\citealt{stalin}. }}
\vfill
\end{minipage}
\end{table*}
   \begin{figure}
   \centering
   \includegraphics[]{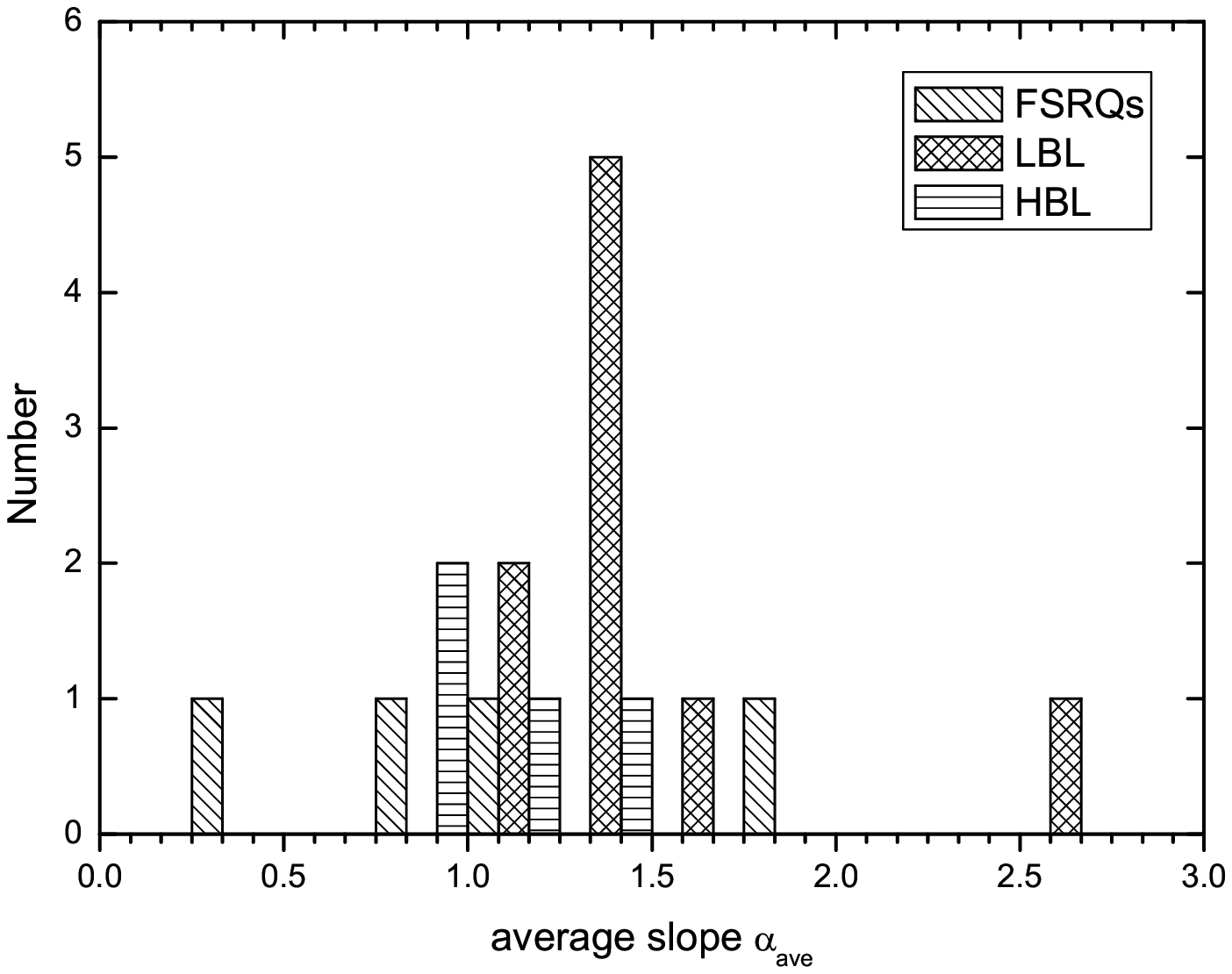}
   \caption{Histogram of the average optical spectral slopes for our 17 samples by subclass }
              \label{ourdistribution}%
    \end{figure}

   \begin{figure}
   \centering
   \includegraphics[]{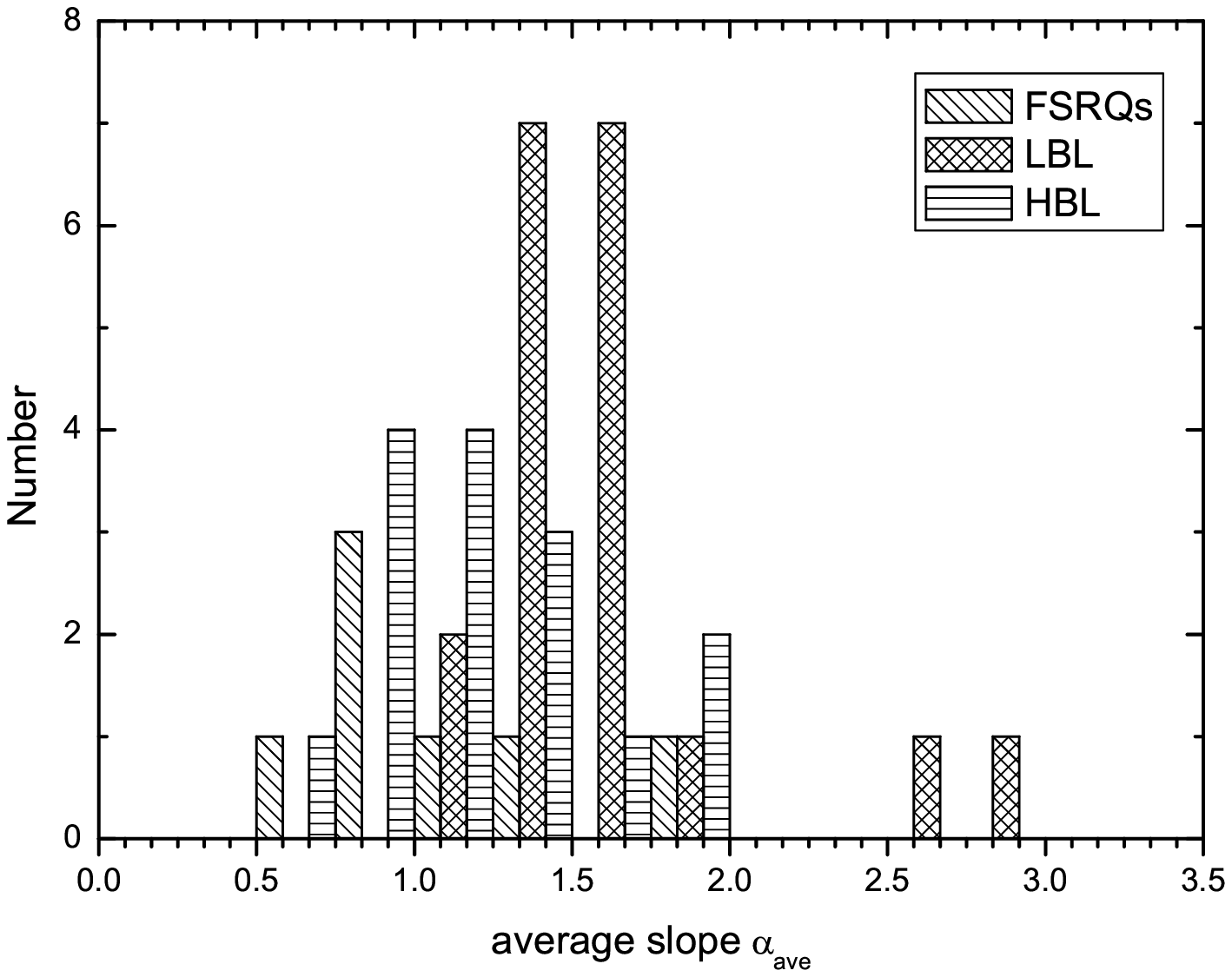}
   \caption{Histogram of the average optical spectral slopes for 41 samples by subclass }
              \label{alldistribution}%
    \end{figure}

\section{Average optical spectral slope distribution}
Blazars are classified into three subclasses
\citep{donato,padovani}. This classification is based upon the
physical natures and characters. So physical connections among
these three subclasses should exist in many fields. Many
investigators studied such connections
\citep{giommi1990,lamer,stecker,delia}, which would play an
important role in constraining the radiation mechanism of blazars
and understanding the fundamental nature of blazars.
\citet{fossati} showed that blazars form a sequence, the SED
changed continuously with the change of their bolometric luminosity. The
synchrotron component and inverse Compton component of HBLs
approximately had the same scale of energy and showed two emission
peaks, which shifted to lower frequencies as the bolometric
luminosity increases. Subclasses from FSRQs through LBLs to HBLs
exhibited a sequence of increasing spectral hardness with the
decreasing of luminosity \citep{bottcher2002}. It suggests that
this may be a sequence of decreasing accretion rate along the same
sequence, so they proposed an evolutionary scenario linking FSRQs,
LBLs and HBLs through the gradual depletion of the circumnuclear
environment of a supermassive black hole. This trend provides the
opportunity to unite three subclasses of blazars under one single
scheme.

Analyzing the dereddened optical spectrum can distinguish the
optical radiation component of three subclasses of blazars
properly. The average optical spectral slopes are listed in cloumn
4 of Table~\ref{table1}. Our results are consistent with the
results of \citet{damicis} for 6 common LBLs within 1
$\sigma$(standard deviation), and they are in good agreement with the
results of \citet{fiorucci} for 14 common blazars within 1
$\sigma$, but the adopted photometric observations are
different. The distribution of $\alpha_{ave}$ for our 17 samples
is given in Fig.~\ref{ourdistribution} by subclass. The average of
$\alpha_{ave}$ is 1.07, 1.49 and 1.08 for FSRQs, LBLs and HBLs,
respectively. We give the distribution of $\alpha_{ave}$ in
Fig.~\ref{alldistribution} for 41 blazars by subclass, which
include our 17 samples, 8 samples from \citet{damicis}, 1 FSRQ
from \citet{ramirez}, 37 objects from \citet{fiorucci}. We averaged
the spectral slopes for all the same available sources. The
average of $\alpha_{ave}$ is 1.10, 1.60 and 1.18 for FSRQs, LBLs
and HBLs, respectively. The spectra of LBLs are obviously steeper
than those of HBLs and FSRQs. According to the peak frequency of
synchrotron emission, we expect $\alpha_{ave}>1$ for LBLs and
FSRQs due to their SED in the descending part, but we expect
$\alpha_{ave}\leq1$ for HBLs because of their SED in the ascending
part. The model of \citet{chiang} predicted that for a broad range
of particle injection distribution, Synchrotron Self Compton
(SSC)-loss-dominated synchrotron emission exhibited a spectrum
whose spectral slope was 1.5 in optical band. The histogram in
Fig.~\ref{alldistribution} shows that all $\alpha_{ave}$ of LBLs
are greater than 1 and the distribution peak is around 1.5
(average is 1.60). Our result is consistent with \citet{fiorucci},
it indicates that SSC-loss-dominated mechanism is probably the
main radiation mechanism for LBLs. $\alpha_{ave}$ of HBLs
distribute in the range from 0.5 to 2, its distribution peak is
between 0.75 and 1.25, while the predicted spectral slope under
pure synchrotron emission should be less than or equal to 1. This
means that the optical radiation of HBLs may contain other
components, such as thermal contribution of accretion disk or host
galaxy, nonthermal emission originating from different regions or
different sizes of the relativistic jet. For FSRQs, the
distribution range of $\alpha_{ave}$ is $0.5\sim2$, whose peak is
between $0.75\sim1$, so their optical emission is likely to be
deformed by thermal emission or other components. Synchrotron
component does not dominate and other mechanism can not be
negligible. ``Thermal Bump" (TB) may be an example, \citet{turler}
showed the ``UV bump" for 3C 273. More simultaneous multi-band
observations of larger samples are important to thoroughly study
the emission components of blazars.

   \begin{figure*}
   \centering
   \includegraphics[]{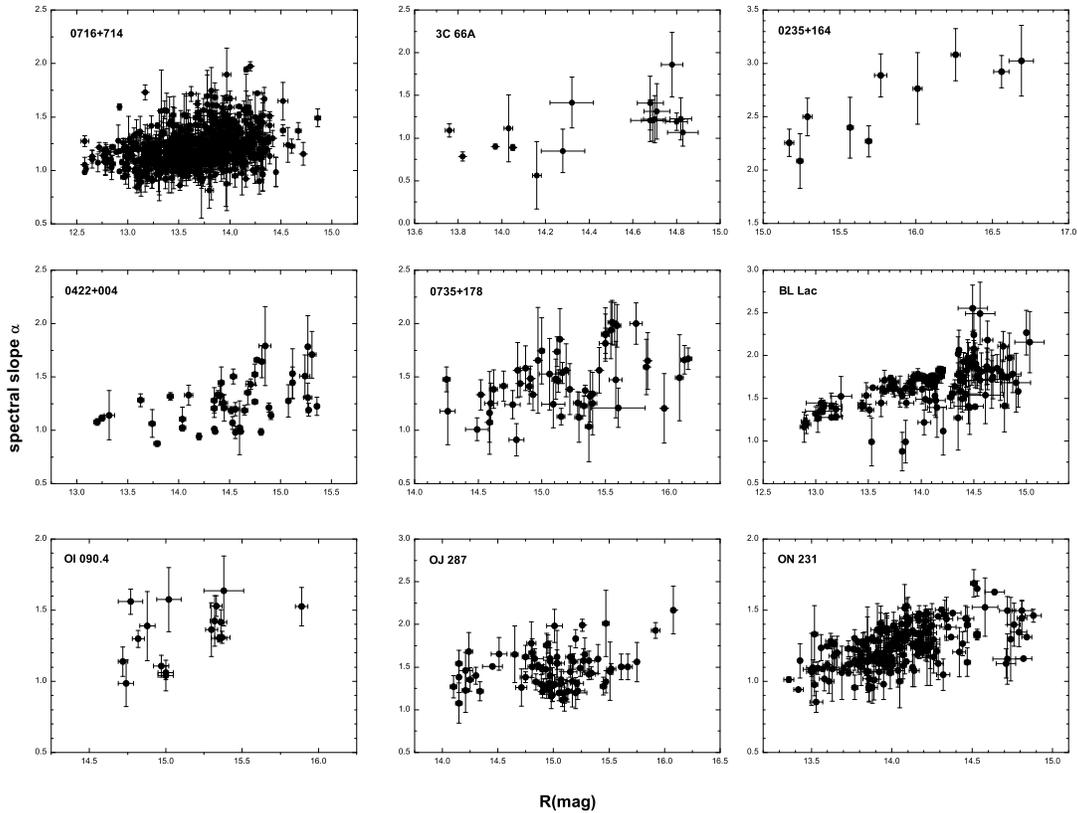}
      \caption{The optical spectral slope versus R magnitude for 9 LBLs}
         \label{lbl}
   \end{figure*}

   \begin{figure}
   \centering
   \includegraphics[]{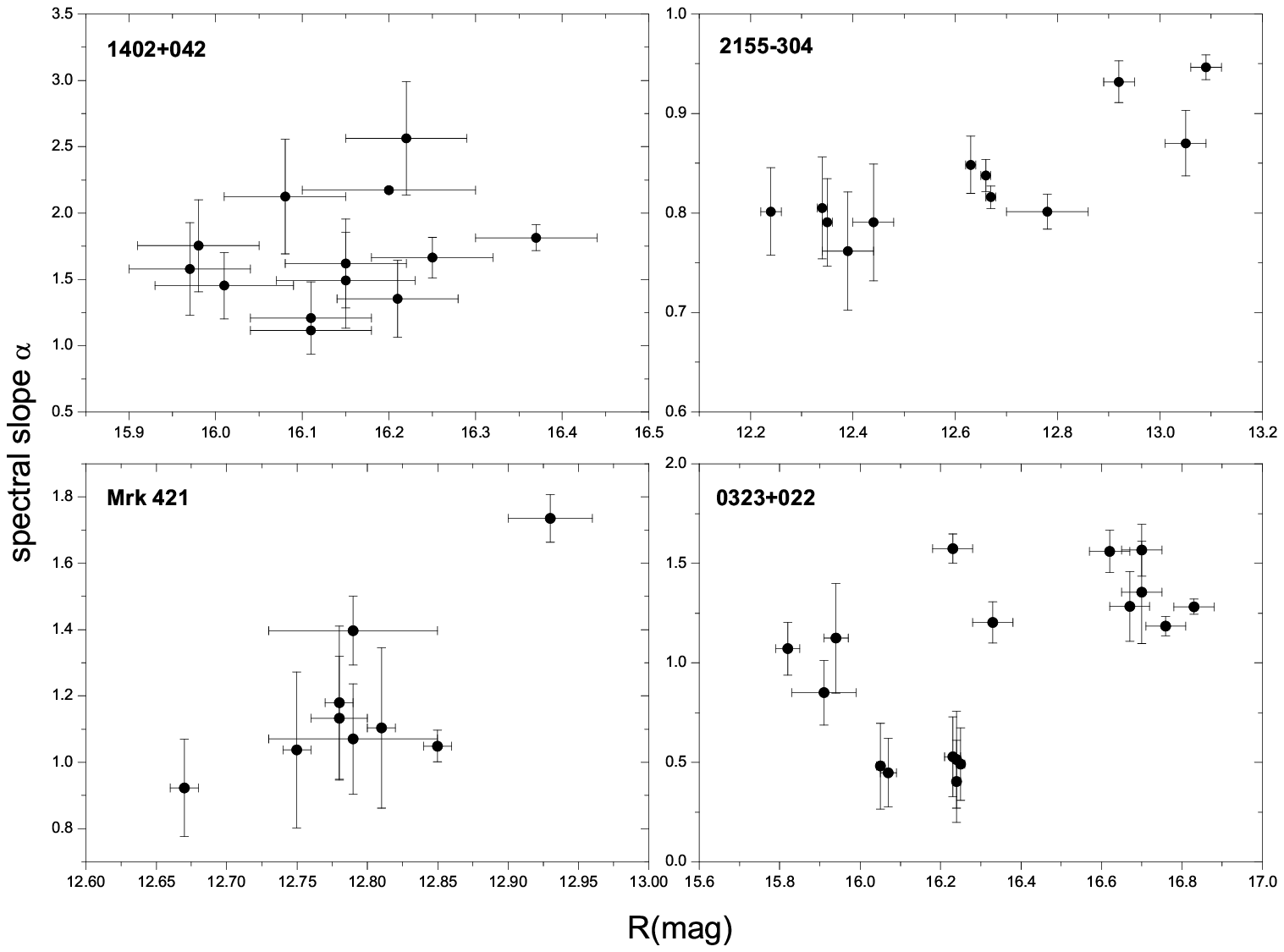}
      \caption{The optical spectral slope versus R magnitude for 4 HBLs}
         \label{hbl}
   \end{figure}

   \begin{figure}
   \centering
   \includegraphics[]{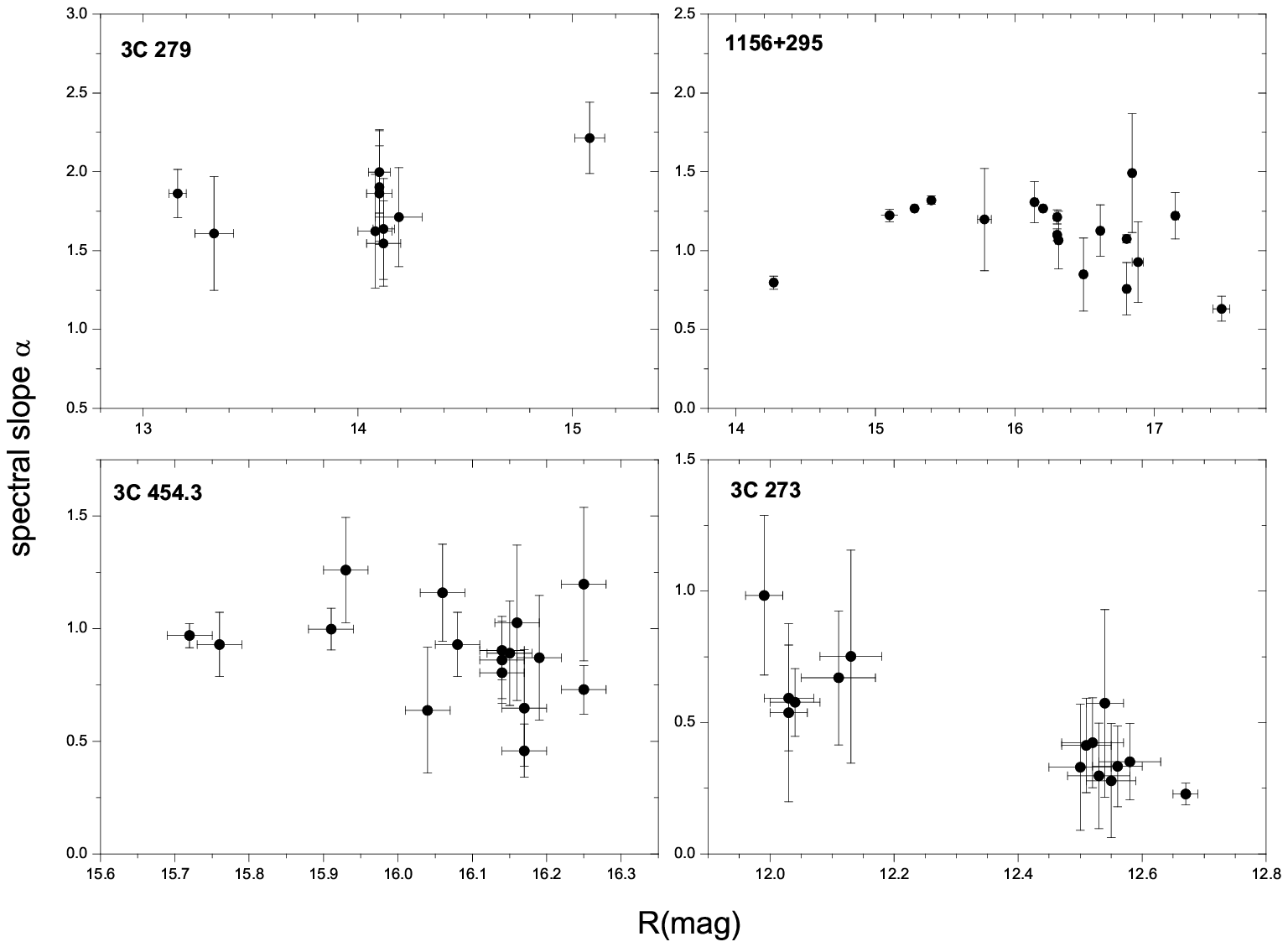}
   \caption{The optical spectral slope versus R magnitude for 4
      FSRQs}
         \label{fsrqs}
   \end{figure}

\section{Spectral slope variability}

The spectral slopes and their uncertainties were obtained by linear
least square fitting with the ``simultaneous" raw data groups.
Figs.~\ref{lbl}$\sim$\ref{fsrqs} show the variability relations
between the spectral slopes and the \emph{R} magnitudes
\citep{massaro1996a} for 9 LBLs, 4 HBLs and 4 FSRQs, respectively.
Linear regression analysis was applied to every sample in order to
investigate the possible relation between the optical spectral
slope and the source luminosity. The linear regression slopes \emph{b},
their standard deviations and the correlation coefficients $r^{2}$
are listed in Table~\ref{table1}. A tendency to become flat with
increasing of \emph{R} magnitude can be observed for all LBLs and HBLs.
$r^{2}$ are greater than 0.5 for 3 out of 4 HBLs and for 5 out of
9 LBLs. The average of correlation coefficient is 0.60 and 0.54
for HBLs and LBLs, respectively. The correlation relations are
significant at a 0.95 confidence level for all HBLs and LBLs
except for 1402+042, the average correlation coefficient of HBLs
is higher than that of LBLs. The variability is consistent with
the relation between color index and luminosity from previous
investigations, and consistent with the variability relation in
other wavebands (e.g.
\citealt{gear,massaro1998,kedziora,papadakis}). Many models
\citep{qian,li,spada,sikora,wang,chiang,vagnetti} stated that
spectra of blazars became flat when they turned bright. For FSRQs,
the spectra generally become steep when they turn bright (e.g.
\citealt{miller,brown1986,clements,ramirez}) except for 3C 279,
but all four correlations are not significant at a 0.95 confidence
level. It is interesting that the tendency is possibly opposite in
the faint state against the bright state for 1156+295. The general
result can be interpreted by the thermal bump model by
\citet{vagnetti}, they predicted that a strong thermal bump could
produce a spectrum that becomes steep when getting bright, where
the thermal bump emerges in the faint state. The variability of
1156+295 is a good support to \citet{brown1989a,brown1989b}, they
mentioned that in high brightness state the spectra
variability of FSRQs showed similar properties with that of BL
Lacs. But in the faint state, since the thermal contribution was
larger in the blue part, the nonthermal component was steeper than
the composite spectrum. The nonthermal component had an even more
dominant contribution to the total flux when the object got
bright, and the composite spectrum got steeper \citep{ramirez}. Our
results suggest that the emission of FSRQs probably contains
thermal component and nonthermal synchrotron component, and the
former one flattens the spectrum.

\begin{figure}
   \centering
   \includegraphics[]{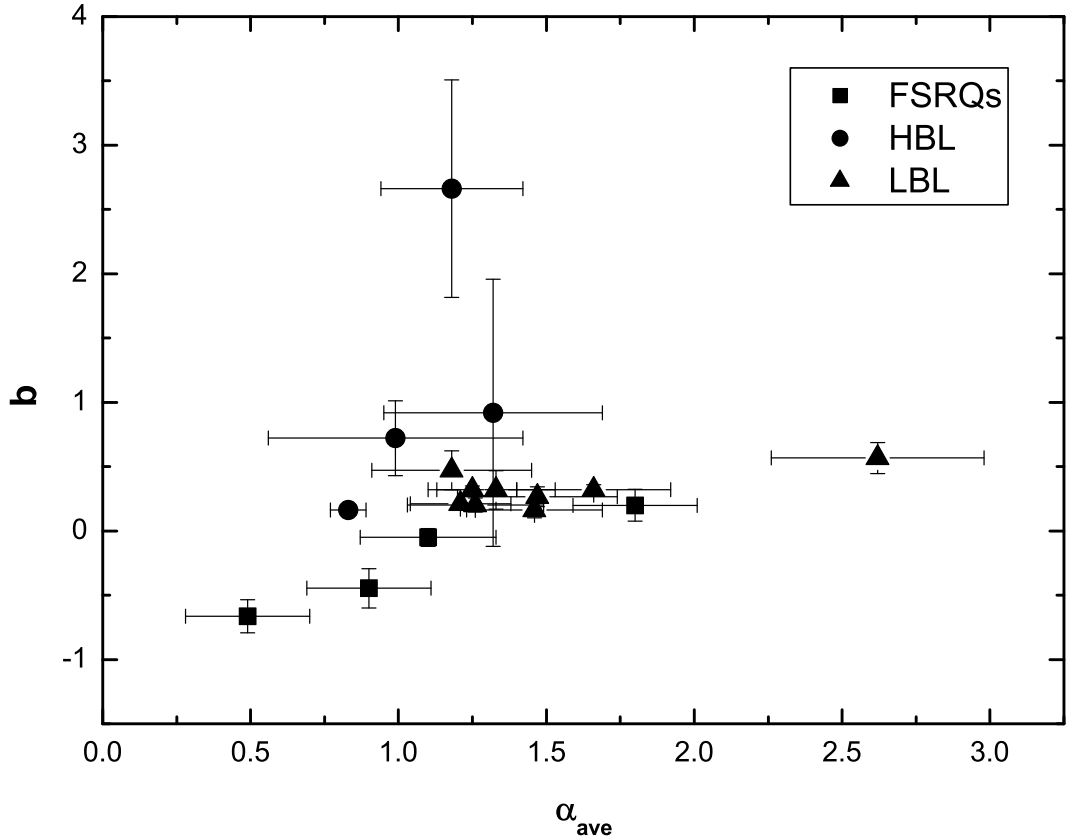}
   \caption{The average optical spectral slope $\alpha_{ave}$ versus the
   slope variability indicator(linear regression slope) b }
              \label{b}%
    \end{figure}

The relation between the regression slope \emph{b} (it is the optical
spectral slope variability indictor) and $\alpha_{ave}$ is shown
in Fig.~\ref{b}. We can see that different subclasses have the
tendency to locate in different regions in this pattern, LBLs are
between FSRQs and HBLs. This may connect with the evolutionary
scenario obtained by \citet{bottcher2002}. The accretion rate
decreases along the sequence of FSRQs, LBLs and HBLs, due to the
gradual depletion of the circumnuclear environment of a
supermassive black hole. More samples are needed to confirm this
result. The mean value of \emph{b} is $-$0.24, 0.32 and 1.12 for FSRQs,
LBLs and HBLs, respectively. The slope variability indicator of
HBLs is the greatest, but the average variable amplitude of \emph{R}
magnitudes is the smallest and the average variable amplitude of
the spectral slope is smaller than that of LBLs.

\begin{figure}
   \centering
   \includegraphics[]{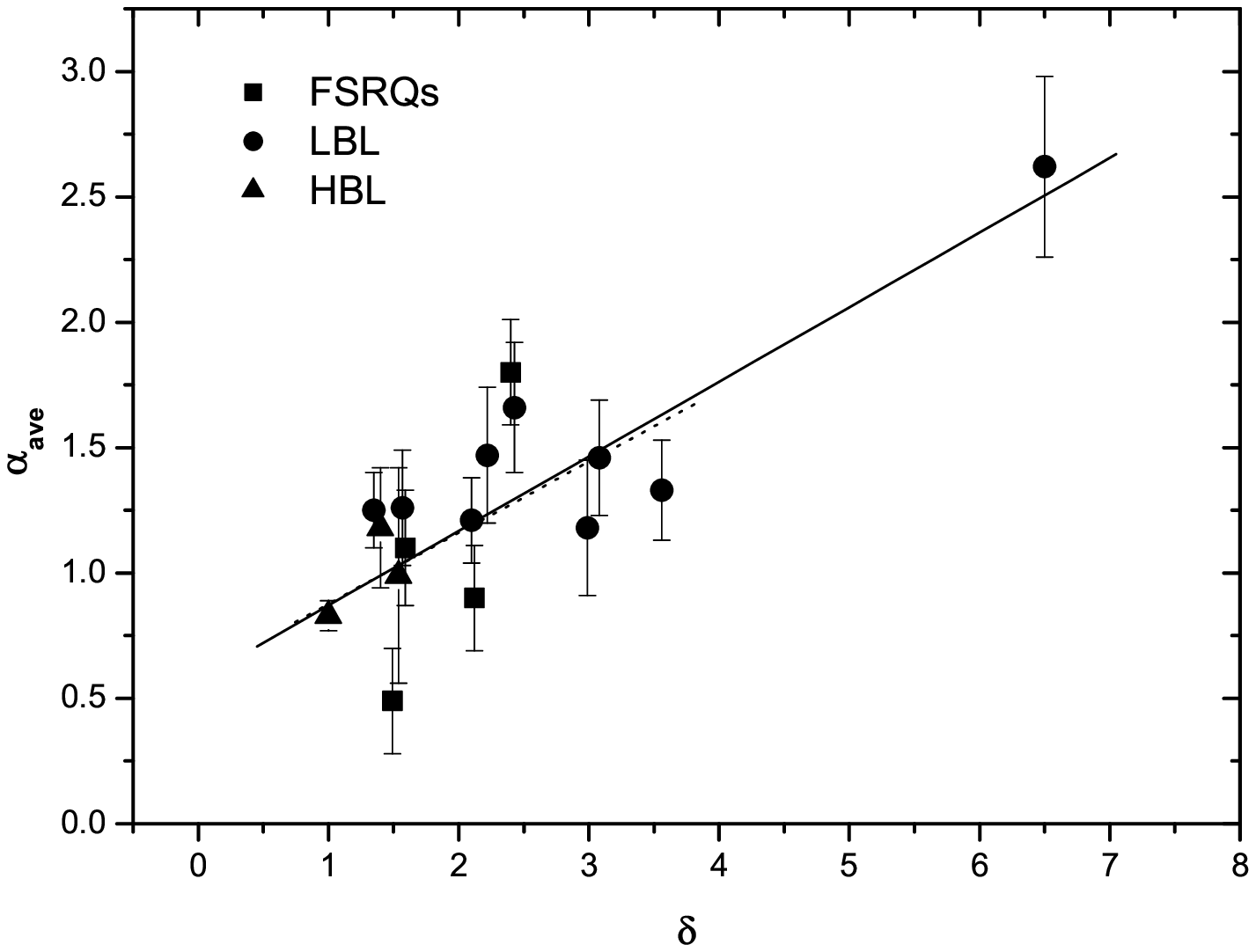}
   \caption{The relation between average optical spectral slope $\alpha_{ave}$ and optical Doppler factor $\delta$.
    Solid line is the linear regression to all the data, and dotted-line is the linear regression to the data except for
    the outlying point 0235+164. }
              \label{doppler}%
    \end{figure}

The observed flux density is enhanced by relativistic Doppler
beaming effect. \citet{ghisellini1989} presented that the jet
plasma might accelerate outward, so that the bulk velocity and the
bulk Lorentz factor increased with increasing distance along the
jet. According to this proposal, \citet{fan1993} and
\citet{xie2001} obtained an empirical formula of
frequency-dependent Doppler factor,
$\delta_{\nu}=\delta^{1+1/8log(\nu_{O}/\nu)}$, where $\nu$ is the
frequency of considered waveband, $\nu_{O}$ is the frequency of
optical band, $\delta$ is the optical Doppler factor,
$\delta_{\nu}$ is the Doppler factor at frequency $\nu$. Therefore
the optical Doppler factor can be obtained if the radio or X-ray
Doppler factor is known. Fig.~\ref{doppler} shows the relation
between $\alpha_{ave}$ and the optical Doppler factor $\delta$.
The linear regression analysis between $\delta$ and $\alpha_{ave}$
shows that the linear correlation is significant at a 0.99
confidence level. The correlation coefficient is 0.79, and the
regression line is plotted with solid line. In order to exclude
the dependence on the outlying point 0235+164, we applied linear
regression analysis to the data excluded 0235+164. The regression
coefficient is consistent with the above-mentioned results within
one standard deviation, the correlation coefficient is 0.70, the
linear correlation is also significant at a 0.99 confidence level.
The regression line is shown with dotted-line in
Fig.~\ref{doppler}. Two regression lines are almost
overlapped, thus the result is not dependent on the outlying point
significantly. The regression equation for all the data is:
\begin{equation}\label{slopedoppler}
    \alpha_{ave}=0.573(\pm0.084)+0.297(\pm0.045)\delta
\end{equation}
We can obtain the optical Doppler factor from average optical
spectral slope if this relation is confirmed. This result suggests
that Doppler beaming effect of relativistic jet is the main
mechanism for the understanding of the properties of blazars.

\section{conclusions}

We collected a large number of ``simultaneous" ($\Delta t<1$
hour and most of their $\Delta t<30$ minutes) optical photometric
data in a long period from 1985 to 2003. Here we gathered a
larger blazar sample which includes some objects from each of
three subclasses. A comparison between these subclasses enables us
to gain some useful information on the emission mechanisms and
emission components. The research of this sample is a great
benefit to investigate the natures of blazars. In this paper we
calculated the long period dereddened optical spectral slopes
by linear least square fitting, which is based on the power law
relation. The average optical spectral slope distributions of
three subclasses of blazars were analyzed. We further on analyzed
the relation between the optical spectral slope and \emph{R} magnitude,
the relation between average optical spectral slope and slope
variability indicator \emph{b}, and the relation between average optical
spectral slope and optical Doppler factor. A discussion mainly
about emission mechanism, emission components and connection of
three subclasses of blazars presents the following conclusions:

   \begin{enumerate}
      \item All the average spectral slopes of our 9 LBLs are greater than
      unity and scatter around 1.5, this is a good support to the SSC-loss-dominated model \citep{chiang}.
      \item The optical spectra of all our HBLs and LBLs become flatter when
      the sources turn brighter, the average linear regression correlation
      coefficient of HBLs is higher than that of LBLs. The relation between spectral slope and \emph{R}
      magnitude of FSRQs suggests that the emission of FSRQs probably has ``thermal bump"
      contribution and the thermal component must be considered.
      \item The slope distribution and variability for HBLs indicate that their spectra
      appear to be deformed by other components, which presumably come from the thermal
      accretion disk or originate from different regions of the jet.
      \item Different subclasses of blazars have the tendency to locate in different
      regions in the pattern of the slope variability indicator versus average spectral slope.
      \item There is a significant correlation (correlation coefficient is 0.79) between
      the average optical spectral slope and optical Doppler factor, which suggests that Doppler beaming
      effect of relativistic jet is the main mechanism for blazars.
   \end{enumerate}

\section*{Acknowledgments}

We are grateful to the referee for valuable comments and detailed
suggestions that have been consulted and adopted to improve this
paper very much. We wish to thank professor Paul J. Wiita for
providing their photometric data. This research has made
employment of the NASA/IPAC Extragalactic Database (NED). This
work has been supported by the National Natural Science Foundation
of China under grant Nos. 10521001 and 10433010.

\bsp \label{lastpage}

\end{document}